\documentclass[letterpaper]{article}
\usepackage{natbib,alife13}

\usepackage{overpic}

\title{Evolutionary Transitions and Top-Down Causation}
\author{Sara Imari Walker$^{1, 2, 3}$, Luis Cisneros$^4$ \and Paul C.W. Davies$^{2, 4}$  \\
\mbox{}\\
$^1$NASA Astrobiology Institute, USA\\
$^2$ BEYOND: Center for Fundamental Concepts in Science, Arizona State University, Tempe AZ USA \\
$^3$ Blue Marble Space Institute of Science, Seattle WA USA\\
$^4$ Center for the Convergence of Physical Science and Cancer Biology, Arizona State University, Tempe AZ USA\\
sara.i.walker@asu.edu}

\begin{document}
\maketitle

\begin{abstract}
Top-down causation has been suggested to occur at all scales of biological organization as a mechanism for explaining the hierarchy of structure and causation in living systems \citep{Campbell, Auletta2008, Davies2006, Davies2012, Ellis2012}. Here we propose that a transition from bottom-up to top-down causation -- mediated by a reversal in the flow of information from lower to higher levels of organization, to that from higher to lower levels of organization -- is a driving force for most major evolutionary transitions. We suggest that many major evolutionary transitions might therefore be marked by a transition in causal structure. We use logistic growth as a toy model for demonstrating how such a transition can drive the emergence of collective behavior in replicative systems. We then outline how this scenario may have played out in those major evolutionary transitions in which new, higher levels of organization emerged, and propose possible methods via which our hypothesis might be tested. 
\end{abstract}
%%%%%%%%%%%%%%%%%%%%%
\section{Introduction}
%%%%%%%%%%%%%%%%%%%%%
The major evolutionary transitions in the history of life on Earth include the transition from non-coded to coded information (the origin of the genetic code), the transition from prokaryotes to eukaryotes, the transition from protists to multicellular organisms, and the transition from primate groups to linguistic communities \citep{ SzathmaryMS1997, Jablonka2006}.  A hallmark of many of these transitions is that entities which had been capable of independent replication prior to the transition can subsequently only replicate as part of a larger reproductive whole \citep{Szathmary1995}. A classic example is the origin of membrane bound organelles within modern eukaryotes, such as the mitochondria, which are believed to have emerged through endosymbiosis with prokaryotes that later lost their autonomy \citep{Lynn1967}. Each such transition is typically viewed as marking a drastic jump in complexity: cells are much more complex than any of their individual constituents ({\it i.e.} genes or proteins), eukaryotes are more complex than prokaryotes, multicellular more complex than unicellular organisms, and human societies more complex than individuals. However, although such a hierarchy is conceptually easy to state, in practice it is difficult to determine what, if any, universal principles underlie such large jumps in biological complexity. 

Szathm\'ary and Maynard Smith have suggested that all major evolutionary transitions involve changes in the way information is stored and transmitted \citep{Szathmary1995}. An example is the origin of epigenetic regulation, whereby heritable states of gene activation lead to a potentially exponential increase in the amount of information that may be transmitted from generation to generation (since a set of $N$ genes, existing in two states - on or off - via epigenetic rearrangements, can have $2^N$ distinct states). Such a vast jump in the potential information content of single cells is believed to have led to a dramatic selective advantage in unicellular populations capable of epigenetic regulation and inheritance \citep{JablonkaLamb1995, Lachmann1996}. The reasoning is straightforward: epigenetic factors permit a single cell line with a given genotype to express many different phenotypes on which natural selection might act, thereby providing a competitive advantage through diversification. Importantly, this innovation was crucial to the later emergence of multicellularity by permitting differentiation of many cell types from a single genomic inventory. However, although epigenetic regulation was likely a necessary precondition for the emergence of multicellular organization (at least in extant lineages), it does not necessitate that such a transition from unicellularity to multicellularity will occur or explain {\it how} it occurs. Plenty of protists are capable of phenotypic differentiation but have never made the transition to true multicellularity, although they may exhibit highly collective and coordinated behaviors (see {\it e.g.} \cite{Nedelcu2004} for a discussion of unicellular, multicellular, and a gamut of intermediate forms within the Volvocalean green algal group). More generally, while it is true that changes in how information is stored and transmitted enable the possibility of new levels of organization to evolve, such innovations are not necessarily a sufficient causative agent to drive the emergence of genuinely new, higher-level, entities. 

Therefore to make progress in understanding the major transitions, a key, and oft understated, distinction must be made between the evolutionary innovations leading up to a major transition that enable higher-levels of organization to emerge, and the mechanism(s) underlying the physical transition itself. In general, the majority of unifying work on major evolutionary transitions has focused on the former perspective by outlining the key steps that enabled a particular transition to occur, such as the innovations in information storage and transmission outlined by Szathmary and Maynard Smith \citep{SzathmaryMS1997}. While these innovations are certainly crucial to our understanding of the historical sequence of evolutionary events surrounding each major transition -- such as the example of the appearance of phenotypic differentiation via epigenetic regulation prior to the emergence of multicellularity cited above -- they tell us little about the underlying mechanisms governing the emergence of genuinely new higher-levels of organization. If there are in fact any universal principles common to all such major jumps in biological complexity, we should expect there to be a common mechanism driving each such transition that is not dependent on a precise series of historical (evolutionary) events. In this paper, we focus on those major evolutionary transitions leading to the emergence of new, higher level entities, which are composed of units that previously reproduced autonomously. We propose that these major transitions, corresponding to major jumps in biological complexity, are associated with {\it information gaining causal efficacy over higher levels of organization.} 

To outline our hypothesis, we first present an introduction to top-down causation in biological systems, and outline how a transition to top-down causation via informational efficacy over new, higher levels of organization might enable the emergence of higher-level entities. We then present a toy model, investigating the onset of non-trivial collective behavior in a globally coupled logistic map lattice, to demonstrate how a reversal in the dominant direction of information flow, from bottom-up to top-down, is correlated with the emergence of collective behavior in replicative systems. A key feature of our analysis is to determine the direction of causal information transfer. We then outline how a transition in causal structure may have driven both the origin of life and the origin of multicellularity, as two representative examples of major evolutionary transitions in which new, higher levels of organization emerged. We conclude with some suggestions about possible methods via which our hypothesis might be tested.

%%%%%%%%%%%%%%%%%%%%%
\section{Informational Efficacy and Top-Down Causation} 
%%%%%%%%%%%%%%%%%%%%%

Biological information is a notoriously difficult concept to define \citep{Kuppers}. This difficultly stems in part from the fact that in living systems the dynamics are coupled to the information content of biological states such that the dynamics of the system change with the states and vice versa \citep{Goldenfeld2011, Davies2012}. This is in marked contrast to the traditional approach to dynamics, where the physical states evolve with time but the dynamical laws remain fixed, or change over much longer time scales. The coupling of states to dynamics is   perhaps most evident for the case of the genome, in which the expressed set of instructions -- {\it i.e.} the relative level of gene expression  -- depends on the state of the system -  {\it i.e.} the composition of the proteome, environmental factors, {\it etc.} - that regulate the switching on and off of individual genes. The result is that the update rules change with time in a manner which is both a function of the current state and the history of the organism \citep{Goldenfeld2011}. This feature of ``dynamical laws changing with states'' \citep{Davies2012}, as far as we know, seems to be unique to biological organization and is a direct result of the peculiar nature of biological information (although speculative examples from cosmology have also been discussed, see {\it e.g.} \citep{Davies2006a}). 

\begin{figure}[t]
\begin{center}
\includegraphics[width=2.5in, height=1in]{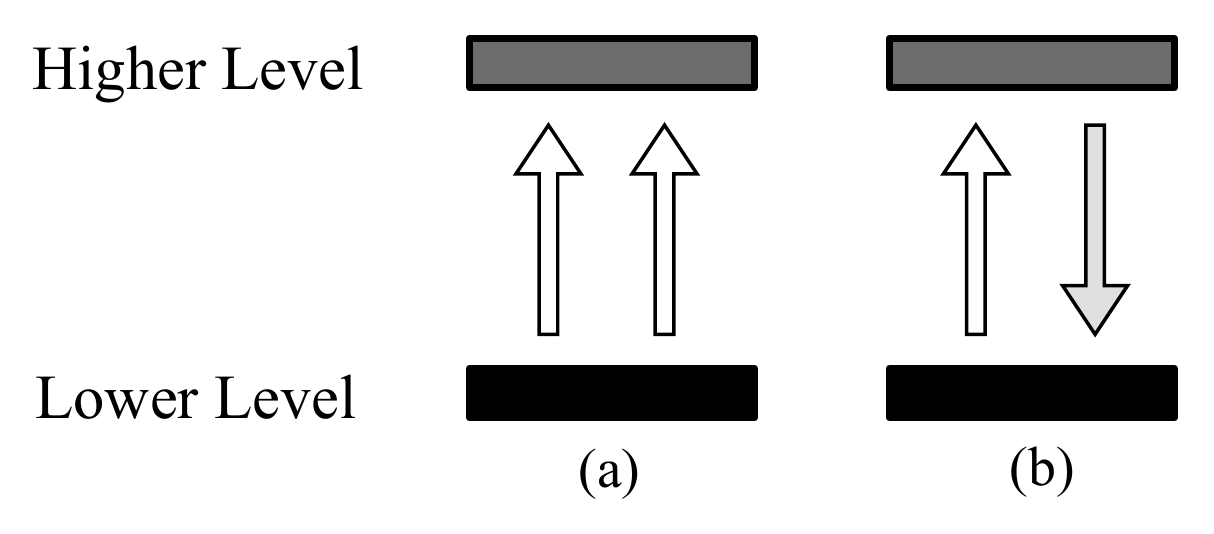}
\caption{Bottom-up and  top-down modes of causation. (a) The standard (reductionist) view suggests everything in the universe is directed by bottom-up action only, such that causation flows strictly from lower to higher levels. (b) Biological organization suggests an alternative causal structure whereby bottom-up modes of causation emanating from lower-levels provide a space of possibilities, and higher-levels of organization modify the causal relations below via top-down causation. Figure adapted from \citep{Auletta2008}.}
\label{fig1}
\end{center}
\end{figure}

Biological information is distinctive in its contextual or semantic nature, in other words it {\it means} something \citep{JMS_2000}. For example, a gene is just a random sequence of nucleotides when taken in isolation, and is indistinguishable from junk, or noncoding, DNA. It is meaningful, or biologically functional,  only within the context of the cell, where a suite of molecular hardware collaborate in decoding and executing the encoded instruction ({\it e.g.} to make a protein). As such, biological information is an abstract global systemic entity, carrying meaning only within the context of an entire living system. It is of course imprinted in biochemical structures, but one cannot point to any specific structure in isolation and say ``aha! I see biological information here!'': even the information in genes is only efficacious and manifested in a relational sense ({\it i.e.} it must be decoded by the appropriate cellular machinery).  Perhaps even more profound, this abstraction appears to have causal efficacy \citep{Auletta2008, Ellis2012, Davies2012} - it is the information that determines the state and hence the dynamics. As such, it is the efficacy of information that leads to the convolution of dynamical laws and states that makes biology so unique.

This convolution results in multidirectional causality with causal influences running both up and down the hierarchy of structure of biological systems ({\it e.g.} both from genome to proteome, {\it and} from proteome to genome via the switching on and off of genes). A full explanatory framework for biological processes should therefore include both bottom-up causation (Fig. 1a) -- such as when a gene is read-out to make a protein that affects cellular behavior -- and top-down causation (Fig. 1b) -- as occurs when changes in the environment initiate an organismal response that permeates all the way down to the level of individual genes \citep{Davies2012}. A striking of the latter is provided by the phenomenon of ``mechanotransduction'', where physical forces, such as the sheer stress on a cell or the YoungÕs modulus of an adjacent surface to which the cell attaches, actually affect gene expression \citep{Alberts}. Bottom-up causation is the status-quo in modern physics, whereas top-down causation is less familiar and difficult to quantify. Generally, top-down causation is characterized by a 'higher' level influencing a 'lower' level by setting a context (for example, by changing some physical constraints) by which the lower level actions take place \citep{Auletta2008, Davies2006, Ellis2006, Ellis2012}. An interesting example of top-down causation is provided by natural selection in evolution \citep{Campbell, Okasha2012}, where the history as well as the fate of an organism is determined by the wider environmental context. This is particularly evident for cases of convergent evolution \citep{Davies2006}, of which the wing is a classic example.  Birds, pterodactyls, and bats each developed wings, despite the fact that their last common ancestor did not possess wings. The commonality of form is attributable to physical (environmental) constraints imposed on wing design, which manifests a particular phenotypic trait in the organism ({\it i.e.} a wing). However, the effect is also a local physical one: the biochemical interactions -- dictated in part by both genetic and epigenetic programming -- that govern the morphological development of something as complex as a bird wing are inherently local. As such, natural selection provides a well-known example of how higher level processes ({\it e.g.} environmental selection) constrain and influence what happens at lower levels ({\it e.g.} biochemistry).\footnote{Although it is normal for biologists to discuss causal narratives in informational terms ({\it e.g.} cells ÒsignalÓ each other, and Òrecruit moleculesÓ to Òexpress instructionsÓ \ldots) a determined reductionist would argue that, in principle, this narrative would parallel an, albeit vastly more complicated, account in terms of molecular interactions alone, in which only material objects enjoy true causal efficacy. In this paper we remain agnostic on the question of such Òpromissory reductionÓ because our principal claims remain valid even if the informational-causal narrative is accepted as a mere {\it fa\c{c}on de parler}.}

\begin{figure}[t]
\begin{center}
\includegraphics[width=2in, height=1in]{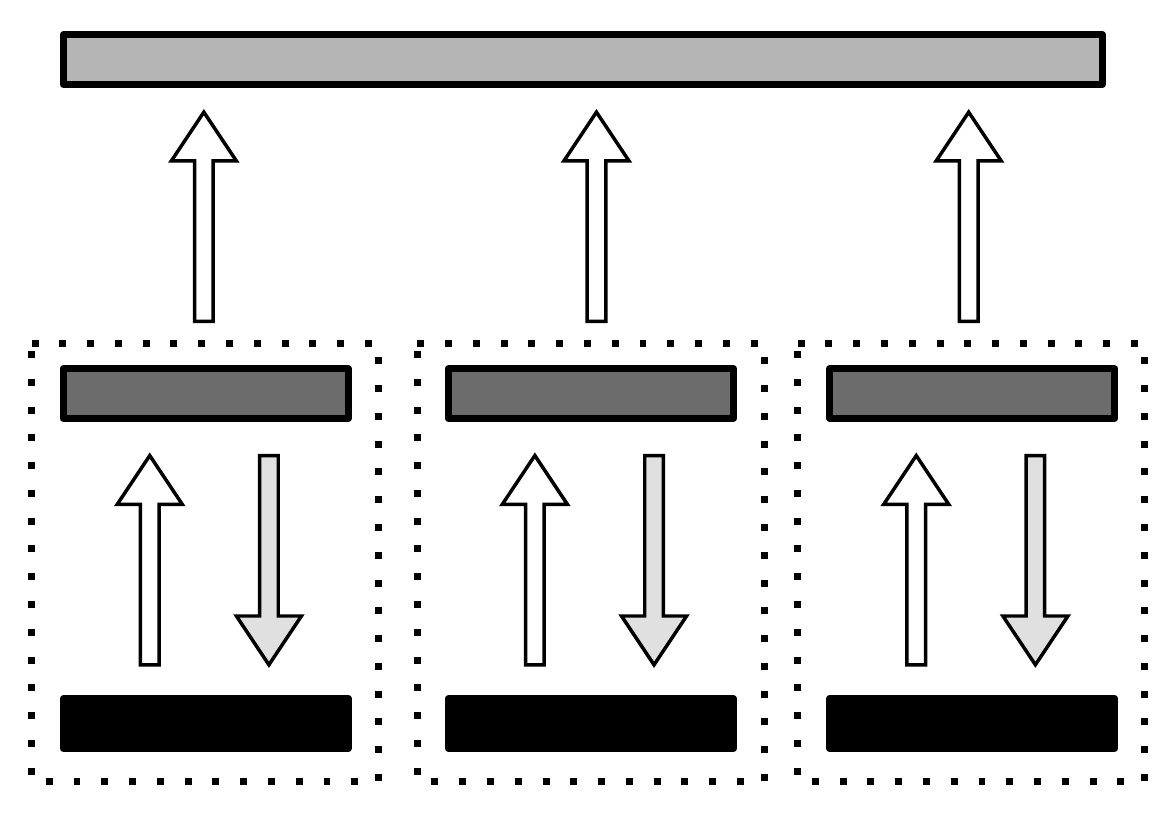}\\
(a)\\
\includegraphics[width=2in, height=1in]{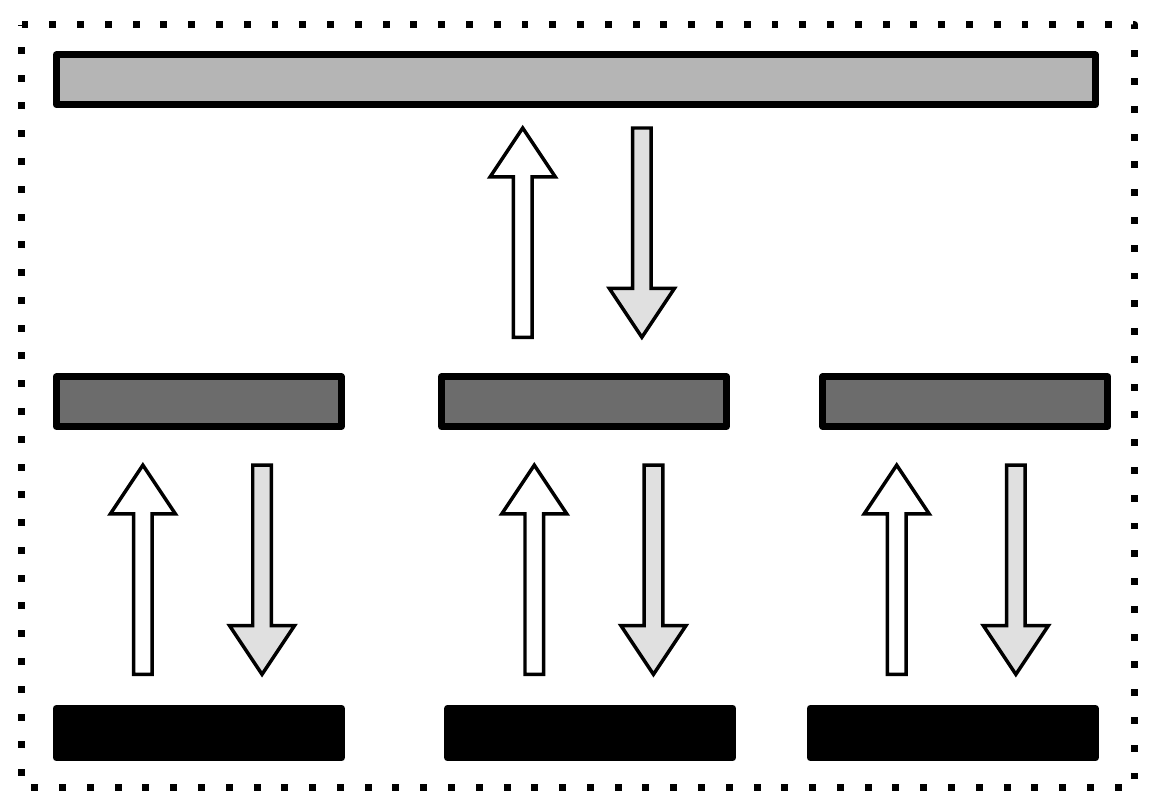}\\
(b)\\
\caption{Schematic illustrating a shift in causal structure mediated by the transition from a collective of lower level entities to a new higher level entity. (a) Prior to the transition, higher levels of organization are dictated by bottom-up causation directed by lower level entities (which themselves may be hierarchal in nature). (b) A new (higher level) entity emerges that may be identified as an ``individual'' when a transition to top-down causation with efficacy over a new, higher-level of organization (in this case the medium grey level) occurs. Dotted lines are used to indicate an individual entity ({\it i.e.} an organism as in the case for the transition from unicellular to multicellular organisms)}
\label{fig2}
\end{center}
\end{figure}

The foregoing discussion indicates that top-down causation -- mediated by informational efficacy -- plays an important role in dictating the dynamics of living systems, where causal influences can run both up and down structural hierarchies. In many of the major evolutionary transitions, new higher-level entities emerge from the collective and coordinated behavior of lower-level entities, eventually transitioning to a state of organization where the lower--level entities no longer have reproductive autonomy. Examples of transitions where lower-level units have lost their autonomy include the the origin of life ({\it i.e.} the hypothetical RNA world), the origin of multicellularity, and the origin of eusociality \citep{SzathmaryMS1997}. During such transitions, the dynamics of lower level entities come under the direction of the emergent high-level entity. This, coupled with the multidirectional causal influences in biological organization, suggests that evolutionary transitions that incorporate new, higher levels of organization into a biological system {\it should be characterized by a transition from bottom-up to top-down causation, mediated by a reversal in the dominant direction of information flow.} Therefore, we suggest that major shifts in biological complexity -- from lower level entities to the emergence of new, higher level entities -- are associated with a physical transition (perhaps akin to a thermodynamic phase transition), and this physical transition is in turn associated with a fundamental change in causal structure (schematically illustrated in Fig. \ref{fig2}). To illustrate this claim, we turn to a toy model investigating the emergence of non--trivial collective behavior in a globally coupled chaotic map lattice \citep{Cisneros2002, Ho2003}.

%%%%%%%%%%%%%%%%%%%%%
\section{Logistic Growth as a Toy Model}
%%%%%%%%%%%%%%%%%%%%%
To explore the emergence of non--trivial collective behavior and its connection to transitions in causal structure, we focus on a lattice of chaotic logistic maps. The logistic growth model was chosen for its connection to the replicative growth of biological populations \citep{Murray1989}, thereby enabling us to make an analogy with the transition from independent replicators to collective reproducers, cited as a hallmark of many major evolutionary transitions as outlined above \citep{Szathmary1995}. Our aim with this simplified model is to provide a clear example of how a reversal in information flow -- from bottom-up to top-down -- can describe a transition from a group of independent low-level entities to the emergence of a new higher-level (collective) entity. 

Our model system is defined as
\begin{eqnarray} \label{eq:coupledmap}
x_{i,n+1} = (1 - \epsilon) f_i (x_{i,n}) + \epsilon~ m_n ~~~;~~~ (i = 1, 2, \ldots N)
\end{eqnarray}
where the function $f_i(x_{i,n})$ specifies the local dynamics of element $i$, $N$ is the total number of elements, $n$ is the current time-step (generation), and $\epsilon$ is the global coupling strength to the instantaneous dynamics of the mean-field, $m_n$, defined below in eq. \ref{eq:mn}. In analogy with biological populations, the element index $i$ may be associated with a specific phenotype within a given population, and $\epsilon$ marks the strength of the global informational control over the local dynamics of each such element. 
The local dynamics of each element $i$ is defined by the discrete logistic growth law
\begin{eqnarray} \label{eq:logistic}
f_i(x_{i,n}) = r_i x_{i,n} \left(1 - \frac{x_{i,n}}{K}\right)  
\end{eqnarray}
where $r_i$ is the reproductive fitness of population $i$, and $K$ is the carrying capacity -- set to $K = 100$ for all $i$, for the results presented here. The instantaneous state of the entire system at time step $n$ is specified as an average over all local states by the instantaneous mean-field $M_n$,
\begin{eqnarray}
M_n = \frac{1}{N} \sum_{j=1}^N x_{j,n}
\end{eqnarray}
and the instantaneous dynamics of the mean-field,
\begin{eqnarray} \label{eq:mn}
m_n =  \frac{1}{N} \sum_{j=1}^N f_j (x_{j,n})
\end{eqnarray}
%It is the instantaneous dynamics of the mean-field, $m_n$, that appears in eq. (\ref{eq:coupledmap}). The instantaneous mean-field $M_n$ is therefore 
is a global systemic entity ({\it i.e.} it cannot be identified with any specific local attribute), which has direct impact on the dynamics of local elements $i$ in our model system. The influence of this abstract global entity is dictated by the global coupling strength $\epsilon$.

The system was initialized with $x_{i,0} = 1$ for all elements $i$, representing an initial population size of one individual for each population. Values for the fitness parameters $r_i$ were randomly drawn from the range of values $[3.9, 4.0]$, where selection of replicative fitness was restricted to this range to ensure that all elements individualistically display chaotic dynamics even when coupled to the global dynamics (required to determine cause and effect for this model system, see {\it e.g.} \cite{Cisneros2002}). Following the dynamics of a set of $N = 1000$ coupled logistic maps, a time series of both the instantaneous states of the local elements, $x_{i,n}$, and of the mean-field, $M_n$, was generated from which causal directionality and the associated flow of information were determined. In what follows, we introduce a definition of a measure of causal information transfer based on analyses of multivariable time series and then present results for the causal structure of our coupled logistic growth model using this measure.

\subsection{Quantifying Causal Information Transfer}
Standard measures of information, such as Shannon entropy \citep{Shannon}, which provides the average number of bits needed to encode independent events of a discrete process, and mutual information, used to measure the joint probability of two process, rely on static probabilities. However, in order to infer causal information transfer ({\it i.e.} from higher to lower levels versus from lower to higher levels of organization, or here from the mean-field to local elements versus from local elements to the mean-field), a measure that can capture {\it dynamical structure} by means of transition probabilities rather than static probabilities is required.
%({\it i.e.} the dynamics must be captured in the relevant transition probabilities)
The  dynamical character of the interactions can be studied by introducing a time lag in order to compute the relevant transition probabilities. 

Consider, for example, a Markov process of order $k$. The conditional probability $
p(x_{n+1}|x_n,\dots,x_{n-k+1}) = p(x_{n+1}|x_n,\dots,x_{n-k+1}, x_{n-k}) $ describes a transition probability whereby each state $x_{n+1}$ of the process is dependent (conditional) on the last $k$-states but is independent of the state $x_{n-k}$ and all previous states. This conditional relationship can be extended to any $k$-dimensional dynamical system as prescribed by Takens embedding theorem \citep{Takens}. To simplify notation, we define an embedded state as $ x_n^{(k)} = (x_n, \dots, x_{n-k+1}) $, which describes a state in the $k$-dimensional phase space, such that the series of vectors $\{x_n^{(k)}\}$ contains all of the information necessary to characterize the trajectory of the dynamical variable $x$. Using this definition, the dynamical information shared between two processes, $x$ and $y$, can be determined by the Transfer Entropy \citep{Schreiber}:
\begin{equation}
T_{Y\rightarrow X} ^{(k)}= \sum_n p(x_{n+1}, x_n^{(k)}, y_n^{(k)})\log\left[\frac{p(x_{n+1}|x_n^{(k)}, y_n^{(k)})}{p(x_{n+1}| x_n^{(k)})}\right] \\
\label{TYX}
\end{equation}
%if all the previous $k$-states are known. Transfer entropy 
This measure incorporates causal relationships by relating delayed (embedded) states, $x_n^{(k)}$ and $y_n^{(k)}$, to the state $x_{n+1}$, and quantifies the incorrectness of assuming independence between the two processes $x$ and $y$. In short, the transfer entropy tells us the deviation from the expected entropy of two completely independent processes. 

The transition probabilities can be systematically measured from the time series by coarse graining the phase space. Calculation of the conditional probabilities
\begin{equation}
p_r(x_{n+1}|x_n^{(k)}, y_n^{(k)})=\frac{p_r(x_{n+1},x_n^{(k)}, y_n^{(k)})}{p_r(x_n^{(k)}, y_n^{(k)})}
\end{equation}
\begin{equation}
p_r(x_{n+1}|x_n^{(k)})=\frac{p_r(x_{n+1},x_n^{(k)})}{p_r(x_n^{(k)})}~.
\end{equation}
then yields all of the necessary quantities required to calculate the transfer entropy $T_{Y\rightarrow X} ^{(k)}$ as defined in equation (\ref{TYX}). Larger values for the information transfer are expected to be measured when the defined embedded space is a better representation of the real phase space of the dynamical process that generates the set of states $\{x_n\}$. Therefore, selection of the dimension of embedding $k$ is done such that $ T_{Y\rightarrow X} = \mathrm{Max}\{T_{Y\rightarrow X} ^{(k)}\}$.

\subsection{Information Flow Between Global and Local Scales}

\begin{figure}[t]
\begin{center}
\begin{overpic}[width=3.25in]%
{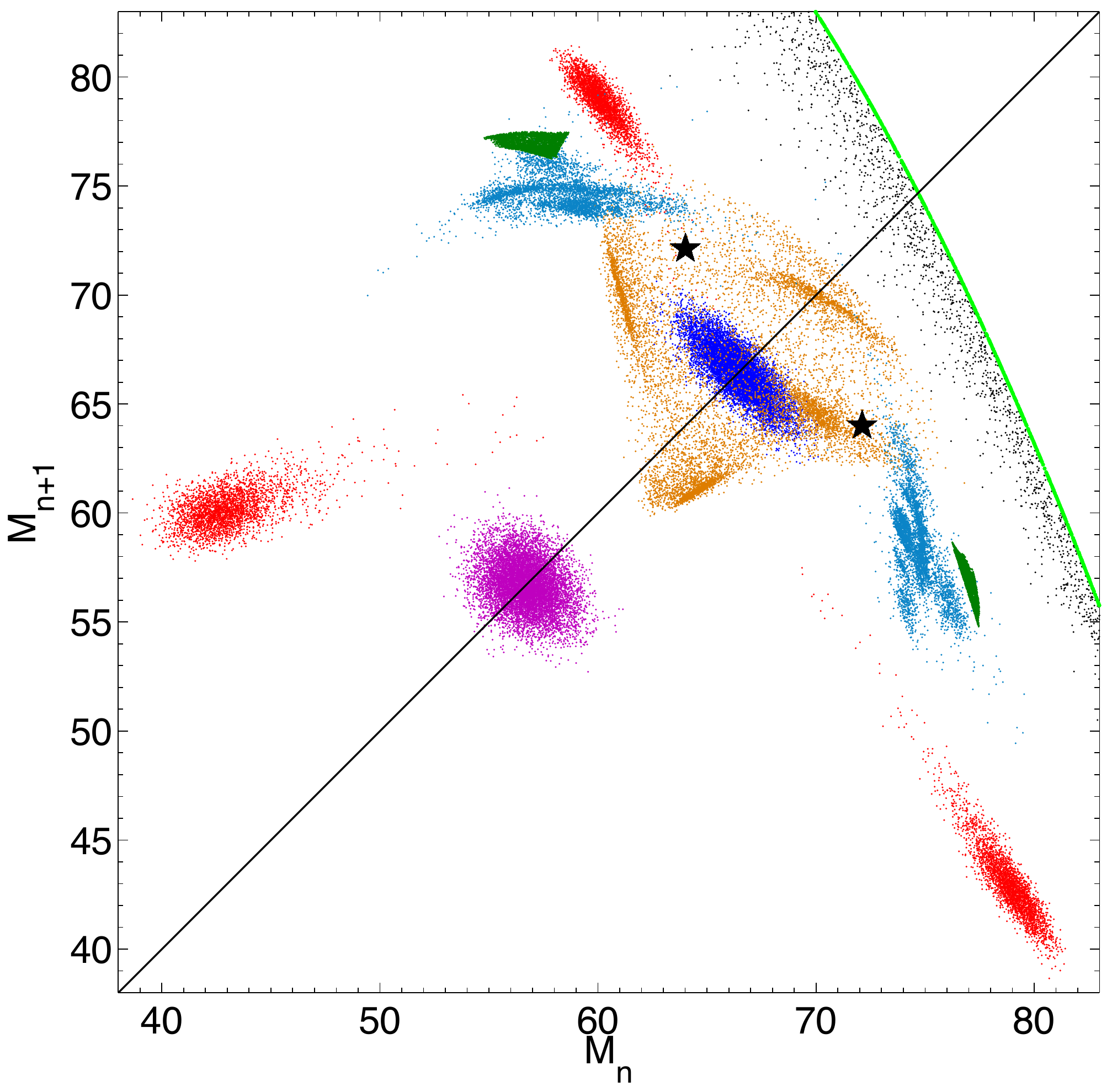}
\put(11,10){\includegraphics[width=1.1in]%
{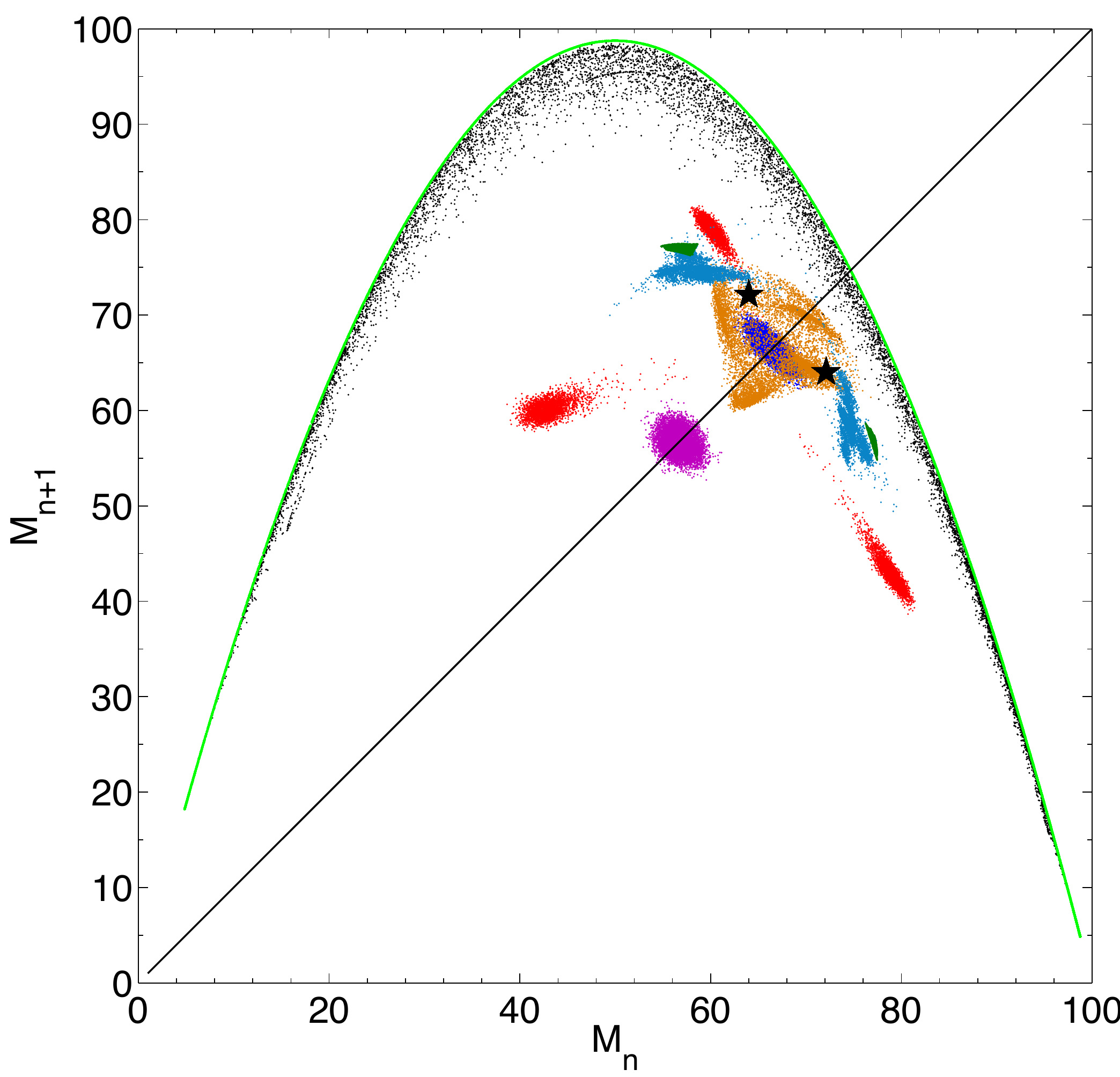}}
\end{overpic}
\caption{Return map for varying values of the global coupling strength $\epsilon$. Shown are return maps for $\epsilon = 0$ (magenta), $\epsilon = 0.075$ (red), $\epsilon = 0.1$ (blue), $\epsilon = 0.2$ (orange), $\epsilon = 0.225$ (aqua), $\epsilon = 0.25$ (dark green), $\epsilon = 0.3$ (stars), and $\epsilon = 0.4$ (black). Also shown is the return map for a single logistic map (bright green). The inset shows an expanded view.}
\label{fig3}
\end{center}
\end{figure}

The coupled system described by eqs. (\ref{eq:coupledmap}), (\ref{eq:logistic}), and (\ref{eq:mn}) displays several different phases with interesting dynamical properties (Fig. \ref{fig3}). A detailed description of the dynamical features of these phases is outlined in the study by Balmforth {\it et al.}  \citep{Balmforth1999}. Here we focus our discussion on the observed collective behavior in the context of the measured causal information transfer characterized in eq. (\ref{TYX}). We compare the flow of information from local to global scales and from global to local scales -- $T_{X \rightarrow M}$ and $T_{M \rightarrow X}$ respectively -- to demonstrate how causal information transfer from the global to the local dynamics corresponds to the emergence of collective organization. 

The time series for $1,000$ logistic maps were recorded for ten thousand generations (time-steps), including the time series of the mean field, $M$, and that of an arbitrarily chosen local element, $x$, selected at random to be representative of typical local dynamical behavior. The dynamics of the system varies widely as a function of the coupling parameter $\epsilon$, indicative of variations in the degree to which the local and mean-field dynamics influence the dynamics of individual local elements. For $\epsilon=0$, the system is completely uncoupled (each local element acts independently), and the dynamics are that of $1,000$ isolated subpopulations. The opposite extreme, $\epsilon=1$, corresponds to complete coupling, where all the logistic maps evolve identically to each other with fully synchronized dynamics ({\it i.e.} they may be identified as part the same higher-level ``organisms''). 

\begin{figure}[t]
\begin{center}
\includegraphics[width=3.25in]{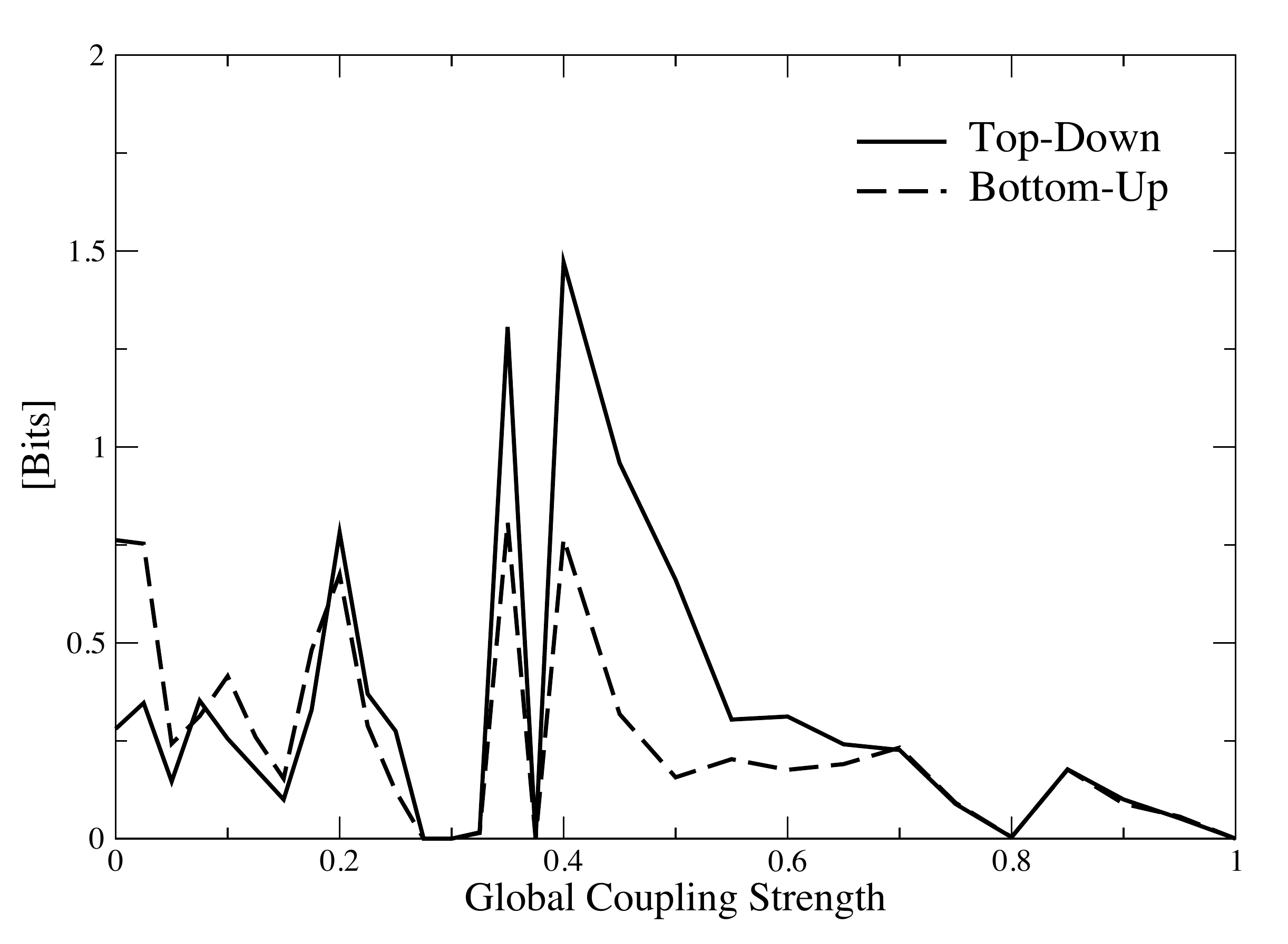}
\caption{Top-down, $T_{M \rightarrow X}$ (solid), and bottom-up $T_{X \rightarrow M}$ (dashed), causal information transfer for varying global coupling strength $\epsilon$ of a system of coupled logistic maps.}
\label{fig4}
\end{center}
\end{figure}

As the coupling strength is increased from no coupling at $\epsilon = 0$, a rich diversity of self-organized collective phenomena are observed to emerge. A sampling of this variety are detailed by the return-map of the mean field $M$, shown in Fig. \ref{fig3}. For $\epsilon = 0$ (the uncoupled limit), the return-map of the mean-field is a cloud of dispersed points around a fixed value (Fig. \ref{fig3}, magenta), as is characteristic of dynamics with random oscillations about a fixed value.  For $\epsilon=0.075$ (Fig. \ref{fig3}, red) a clear quasi-periodic three state oscillatory dynamic is observed, as evidenced by the three clouds in the return map (with some dispersion), indicating that the system has achieved a moderate degree of collectivity. Although the coupling strength is relatively low, the system self--organizes in such a way that the mean field has a simpler dynamic than the typical chaotic behavior of the individuals. The system organizes by forming clusters, within which the individuals have very similar behavior. Here it is likely that top-down information transfer is highest within clusters, resulting in intermediate size scales (between local and global) driving the emergence of collective behavior: this dynamic is not accurately captured by our global measure $M$. As such, the transfer entropy, shown in Fig. \ref{fig4} for top-down ($T_{M \rightarrow X}$, solid) and bottom-up ($T_{X \rightarrow M}$, dashed), do not quite reflect the onset of this collective dynamic, although  $T_{X \rightarrow M}$ and $T_{M \rightarrow X}$ are nearly equal, indicating that the dynamics of the global mean-field is driving, at least partially, the collective behavior in a top-down manner.  

Increasing the coupling strength to $\epsilon = 0.1$, the system falls back into the dynamics of a seemingly disorganized system, and no particularly interesting global behavior is observed.  This is shown in the return map as a randomly dispersed cloud around a fixed value (Fig. \ref{fig3}, blue). However, further increasing the coupling strength to $\epsilon = 0.2$, yields the onset of a new collective phase (Fig. \ref{fig3}, orange). Here, collective behavior manifests as four phase periodic oscillation with large dispersion. A small dominance of the top-down transfer entropy $T_{M \rightarrow X}$ is observed.
% with a corresponding drop in the bottom-up transfer entropy, $T_{X \rightarrow M}$. 
Here, the onset of collective behavior corresponds to a transition to top-down information flow being the dominate mechanism of information transfer. This provides the first presented example of a clear case where top-down causation drives the emergence of collective behavior. It is interesting that this occurs at a fairly low value of the coupling strength, at $\epsilon = 0.2$. 

The attractor observed for $\epsilon  = 0.2$ breaks apart into two dispersed clouds for $\epsilon = 0.225$ (Fig. \ref{fig3}, aqua) and then concentrates into smaller clouds for $\epsilon = 0.25$ (Fig. \ref{fig3}, dark green), where the system enters a collective two-phase periodic oscillation. These observed collective phases also correspond to the top-down transfer entropy being the dominant causal driving force, as shown in Fig. \ref{fig4}. Increasing $\epsilon$ to  $\epsilon = 0.3$ leads to complete synchronization of the system (Fig. \ref{fig3}, stars), yielding a transfer entropy measure of zero. In general, we expect both $T_{M \rightarrow X}$ and $T_{X \rightarrow M}$ to be zero for states of complete synchronization since no information can be gained from a coding by considering a second time series that is dynamically identical to the first one. In other words, when full synchronization is achieved, two series become dynamically identical and it no longer makes sense to discuss transfer entropy between them. In this particular case, full synchronization indicates the local dynamics are coincident with the mean-field, {\it i.e.} a transition to a fully collective emergent entity has occurred. It is interesting that these dynamics are first observed for such a low value of $\epsilon$, and that increasing $\epsilon$ still further yields states which are not fully synchronized. Additionally, in this regime where synchronization emerges dynamically, any collective dynamics in which transfer entropy can be measured ({\it i.e.} not synchronized) are dominated by top-down transfer entropy ({\it e.g.} see $\epsilon = 0.25$ and $0.35$ in Figure \ref{fig4}), suggesting that the dynamical synchronization occurs due to top-down dynamical driving.
For $\epsilon = 0.4$, the return map approaches the form of the chaotic attractor for a logistic map (Fig \ref{fig3}, black), indicating that the dynamics are collectively logistic. The mean field is dynamically chaotic, as are the individual maps in the lattice; however, here the individual maps are not synchronized with each other, yielding non-trivial organization. The emergence of this highly collective state is again reflected by a dominance of the top-down transfer entropy relative to the bottom-up transfer entropy. This trend is continued for increasing $\epsilon$ until $\epsilon = 0.7$, at which point the mean-field and local dynamics are fully synchronized and it no longer makes sense to discuss information transfer, as noted above. 

In general, the trends observed indicate that each time a collective state emerges, causal information transfer is dominated by information flow from global to local scales. Particularly interesting is that top-down causation dominates for collective states in regimes with $0.2 <  \epsilon < 0.7$, where the contribution from the global dynamics is not necessarily the dominant contribution in eq. (\ref{eq:coupledmap}) ({\it i.e.} for $0.2 < \epsilon < 0.5$). In this regime, although the weight of the contribution from the global scale may be less than the contribution from the local scale in dictating the local dynamics, collective states self-organize which are driven by top-down causal information transfer from the mean-field. Although we have focused on a coupling to the global mean-field for the work presented here, other studies of coupled chaotic map lattices have shown that strictly local coupling leads to similar dynamical behavior \citep{Cisneros2002, Ho2003} - {\it i.e.} even in cases where the mean-field never appears in the dynamical equations, the global dynamics can still drive the emergence of collective behavior via top-down causation.

%%%%%%%%%%%%%%%%%%%%%
\section{Major Transitions in Causal Structure} \label{MT}
%%%%%%%%%%%%%%%%%%%%%

The results presented for this toy model system indicate that a transition from a population of independent replicators, to a collective representing a higher-level of organization, can be mediated by a physical transition from bottom-up to top-down information flow, where non-trivial collective behavior is associated with the degree to which local elements receive information from the global network. The dynamical system investigated was designed to parallel transitory dynamics believed to be a hallmark feature of many major evolutionary transitions -- {\it i.e.} those characterized by the emergence of higher--level reproducers from lower level units \citep{Szathmary1995}. For the model system presented above,  new high-level entities would be expected to emerge as $\epsilon \rightarrow 1$ (although non--trivial collective behavior is observed to emerge in intermediate regimes, as discussed above). Examples of major evolutionary transitions where similar dynamics are expected to have played out include the origin of life, the origin of eukaryotes, the origin of multicellularity, and the origin of eusociality. Here we focus on discussing the origin of life and the origin of multicellularity as two representative examples of major evolutionary transitions that may potentially be driven by transitions in causal structure as dictated by informational gaining efficacy over higher-levels of organization. 

\paragraph{The Origin of Life.}

In the original classification scheme of Szathm\'ary and Maynard Smith, three major transitions are associated with the origin of life: from replicating molecules to populations of molecules in compartments, from unlinked replicators to chromosomes, and from RNA to RNA + DNA + protein ({\it i.e.} the origin of the genetic code). However, given that we do not know the specific sequence of events leading to the emergence of the first known life, a more pragmatic perspective is to 
assume that when life as we know it first emerged, it was surely characterized by the same distinctive hierarchical and causal structure as all known life.  Adopting this viewpoint, Walker and Davies have recently suggested that a transition in causal structure, from bottom-up to top-down, was the critical step in the origin of life \citep{Walker2012}. In this context, the origin of life is associated with the emergence of a collective contextual information processing system with top-down causal efficacy over the matter it is instantiated in \citep{Walker2012}. The transition from non-living to living matter may therefore be identified when information (stored in the state of the system) gains causally efficacy. A constructive measure of how close chemical systems are to the living state -- a quantity notoriously absent in almost all discussions of the origin of life -- may therefore be provided by adopting a variant of the parameter $\epsilon$ and applying it to the relevant chemical kinetics. This may provide new avenues of research into the origin of life by directing efforts toward understanding how chemical systems come under direction of the global context rather than focusing strictly on the evolutionary processes that might enable a transition to the living state but do not necessitate it.     

\paragraph{The Origin of Multicellularity.}

Unlike the emergence of life, where the frequency of origination events is entirely unknown, multicellularity is believed to have arisen dozens of times in the history of life on Earth \citep{Bonner1999}. A possible explanation for the numerous transitions to multicellularity is that many of the hallmarks of multicellular organisms are laid out by epigenetic factors and physical effects in unicellular aggregates that only later come under information ({\it i.e.} genetic) control. For example, Newman and collaborators have proposed that the variety of metazoan body plans were originally laid out by physical interactions, such that the phenotype of multi-cellular aggregates was determined at first by physical environmental influences \citep{Newman2000, Newman2006}. They suggest that these physical varieties of form were only later to be taken over by innovations in genetic programming. An explicit example of a similar process whereby information control dictates the emergence of collective states is provided within the genus Volvox: the multicellular green alga, {\it Volvox carteri} has a gene controlling cellular differentiation that is related to an analogous gene dictating cellular phenotype in its unicellular relative, {\it Chlamydomonas reinhardtii} \citep{Nedelcu2009}, which may have played a crucial role in its transition to multicellularity. This suggests that a key feature of the transition to multicellular organization is biological information gaining efficacy over new scales of organization by redirecting features already present in collectives of the lower-level units. As such, the physical transition should be marked by a transition in causal structure. An interesting consideration is therefore that multicellularity emerges frequently, requiring only the physical transition from bottom-up to top-down causation via information control once the underlying lower-level units posses evolutionary innovations necessary to prime them for the transition. An important question in then: how hard is it for the physical transition to occur? From the viewpoint of the perspective provided here, further investigations into the causal structure of biological systems are required to address this question. The relevant order parameter $\epsilon$ could be measure of the degree of signaling between individual cells ({\it i.e.} their response to intercellular signaling), or a measure reproductive viability as presented with the simple logistic growth model detailed above. In general this approach requires innovations in understanding the degree to which the whole dictates the parts in biological collectives, as much as understanding the degree to which the parts dictate the whole.   

Given that the we do not have a clear picture of the causal structure of biological systems, it is at present unclear what the relative role of bottom-up and top-down causative effects are in directing biological organization. Here we have proposed that increasing levels of biological complexity, corresponding to increased depth in the hierarchical organization of living systems, correspond to information gaining causal efficacy over increasingly higher levels of organization. Each major evolutionary transition leading to the emergence of genuinely new, higher-level entities from lower-level units, should therefore be characterized by a transition in causal structure mediated by a reversal in the dominant direction of information flow from bottom-up to top-down. We have demonstrated the dynamics of such a transition by appealing to a toy system of coupled logistic maps. The dynamics observed verify that collective states emerge in association with a transition to top-down causal information transfer as the dominant direction of information flow. The nature of the reversal in causal structure presented here suggests that biological systems cannot jump up the ladder of hierarchical structure - information must first gain control over a lower-level of organization before the emergence of efficacy over higher--levels can take-hold. Rapid diversification may occur after each such transition due to the new capacity for directing physical processes at the higher level.

%%%%%%%%%%%%%%%%%%%%%
\section{Acknowledgements}
%%%%%%%%%%%%%%%%%%%%%
SIW gratefully acknowledges support from the NASA Astrobiology Institute through the NASA Postdoctoral Fellowship Program. LC and PCWD were supported by NIH grant U54 CA143682.

\footnotesize
\bibliographystyle{apalike}
\bibliography{EvoTran}

\end{document}